\documentclass[11pt]{article}

\usepackage{graphicx}
\usepackage{amsmath}
\usepackage{dcolumn}
\usepackage{bm}
\usepackage{slashed}
\usepackage[bookmarks,bookmarksnumbered,colorlinks=true,anchorcolor=blue,
linkcolor=blue,urlcolor=blue,citecolor=blue,breaklinks=true]{hyperref}
\usepackage[utf8]{inputenc}
\usepackage{authblk}
\usepackage{amsmath}
\usepackage{newtxtext,newtxmath}
\usepackage{color}
\usepackage{ulem}

\newcommand{\be}{\begin{equation}}
\newcommand{\ee}{\end{equation}}
\newcommand{\ba}{\begin{eqnarray}}
\newcommand{\ea}{\end{eqnarray}}
\def\bea{\begin{eqnarray}}
\def\eea{\end{eqnarray}}

\newcommand{\gsim}{\mathrel{\hbox{\rlap{\lower.55ex \hbox {$\sim$}}
                   \kern-.3em \raise.4ex \hbox{$>$}}}}
\newcommand{\lsim}{\mathrel{\hbox{\rlap{\lower.55ex \hbox {$\sim$}}
                   \kern-.3em \raise.4ex \hbox{$<$}}}}

\def\roughly#1{\mathrel{\raise.3ex\hbox{$#1$\kern-.75em%
\lower1ex\hbox{$\sim$}}}}
\def\lsim{\roughly<}
\def\gsim{\roughly>}

\def\({\left(}
\def\){\right)}
\def\[{\left[}
\def\]{\right]}
\def\<{\langle}
\def\>{\rangle}

\usepackage{xcolor}

\title{\bf {Geometric entropy and time-like entanglement entropy on a rotating BTZ black hole}}

\author[1]{Huayu Dai\thanks{huayudai@link.cuhk.edu.cn}}
\author[2]{Xi-Hao Fang\thanks{fangxh9@alumni.sysu.edu.cn}}
\author[3]{Mitsutoshi Fujita\thanks{fujitamitsutoshi@usc.edu.cn}}
\author[4]{Song He\thanks{hesong@nbu.edu.cn}}
\affil[1]{School of Science and Engineering, The Chinese University of Hong Kong (Shenzhen),
Shenzhen 518172, China}
\affil[2]{Beijing Institute of Mathematical Sciences and Applications (BIMSA), Huaibei Town, Huairou District, Beijing 101408, China}
\affil[3]{School of Nuclear Science and Technology, University of South China, Hengyang 421001, China}
\affil[4]{Institute of Fundamental Physics and Quantum Technology, \& School of Physical Science and Technology, Ningbo University, Ningbo, Zhejiang 315211, China}

\date{\today}

\begin{document}

\maketitle

\begin{abstract}
In this paper, we analyze the double Wick rotation of a rotating BTZ black hole and the entanglement entropy. We derive the transition matrix dual to the double Wick-rotated BTZ black hole, which has the usual shape at an imaginary chemical potential. In the dual gravity side, the double Wick rotated BTZ black hole, which is obtained as a quotient, is equal to a rotating BTZ black hole after the coordinate transformation and the identification of periodicity. The geometric entropy and time-like entanglement entropy are reproduced by the identification. {New Lorentzian entanglement growth is defined by the coefficient of linear growth of time-like entanglement entropy.}

\end{abstract}
{}

\newpage

\allowdisplaybreaks

\flushbottom

\section{Introduction}
An intriguing discovery was found in the boundary entropy $S_0=\log g$ ($g$ is the ground state degeneracy) of a 1d system that had open boundaries~\cite{Affleck:1991tk, Cardy:1984bb} and in a changed periodicity of spacetime, which resulted in modular transformations in the entanglement entropy~\cite{Azeyanagi:2007bj}. A conformal transformation to the cylinder can also result in a change of periodicity when two-point functions of twisted operators are considered~\cite{Calabrese:2004eu}: The cut was made either in a circle or along the line of a cylinder.

The geometric entropy is the double Wick rotated version of the entanglement entropy, which makes it possible to use twisted operators in $2d$ CFT. Twisted operators are inserted along Euclidean time in correlation functions. Geometric entropy is not dependent on the standard modular Hamiltonian, but the density matrix is calculated through a spatial direction as Euclidean time and its Hamiltonian~\cite{Fujita:2008zv}. This quantity is well-defined on a higher-dimensional compact manifold $S^3$, unlike the Polyakov loop, and is analogous to the thermodynamical entropy for quantum field theory on general Euclidean manifolds. The geometric entropy has been used to study large $N$ phase transitions in Yang-Mills theory~\cite{Fujita:2008zv,Gromov:2014kia} and QCD-like theory with flavor~\cite{Fujita:2010gx} on compact space. However, geometric entropy was not well understood in terms of degrees of freedom in a time-dependent system evolving under a time-dependent density matrix, or in a system dual to non-static backgrounds, such as a rotating BTZ black hole. See also~\cite{Hubeny:2007xt} for the gravity dual of the covariant entanglement entropy and the time-like entanglement entropy in a rotating BTZ black hole.

The double Wick rotation in the gravity dual allows for the analysis of geometric entropy,  while the double Wick-rotated BTZ black hole is not a black hole for real angular momentum and has a closed time-like curve~\cite{Fujita:2022zvz}. It arises behind the Cauchy horizon (a lightlike boundary)~\cite{Banados:1992gq}. It is necessary to demonstrate how the physics laws permit a closed time-like curve~\cite{Friedman:1990xc}. It was pointed out that many physical quantities in the double Wick rotated background agree with those of a rotating BTZ black hole with the same periodicity~\cite{Dai:2025yhi}. The motivation of this paper is to gain insight into this equivalence by utilizing both field theories and the gravity dual.

The purpose of this paper is to analyze geometric entropy and time-like entanglement entropy in a rotating BTZ black hole, in which the CFT side is the thermal CFT at a chemical potential for momentum. Holographic models can be utilized to compute geometric entropy. The geometric entropy was utilized to probe AdS Schwarzschild black holes~\cite{Bah:2008cj} and the Reissner Nordstr{\" o}m AdS background (effects of background charges)~\cite{Allahbakhshi:2013wk} by using a minimal surface. The relationship between geometric entropy and the DBI action was highlighted. The gravity dual of the time-like entanglement entropy is non-trivial due to the complex area, which is a combination of both spacelike and timelike extremal surfaces that are homologous to the subregion~\cite{Doi:2022iyj, Doi:2023zaf}. In 2-dimensions, conformal symmetry is useful, and the double Wick rotation  in the Euclidean signature is just given by changing the complex variable $z=x+i\tau_E$ as $z=-iz'$. Analyzing geometric entropy and the transition matrix is a good beginning point.


This paper is organized in the following manner. In section \ref{SEC1}, the transition matrix that is dual to the double Wick rotated BTZ black hole is derived and analyzed using an imaginary chemical potential.
CFT is used in Section \ref{SEC3} for computing the expected value of the stress energy tensor. In section \ref{SEC4}, the double Wick rotated BTZ black hole is obtained as quotients. The double Wick rotated BTZ black hole is transformed into a rotating BTZ black hole, as demonstrated in section \ref{SEC5} during the coordinate transformation. Covariance is used to derive the geometric entropy. The time-like entanglement entropy is found after the identification, and its time-dependence is resolved in section \ref{SEC6}.

\section{The transition matrix of the double Wick rotated theory}\label{SEC1}
The transition matrix dual of the double Wick-rotated BTZ black hole is obtained in this section, and it is used to calculate the geometric entropy. It is demonstrated that its structure is analogous to that of a rotating BTZ black hole with an imaginary chemical potential. The metric in the gravity dual determines the periodicity. 
The Euclidean version of the rotating BTZ black hole \cite{Banados:1992gq, Banados:1992wn} is the first thing we analyze as follows:
\ba\label{BTZ1}
ds^2=l^2\Big(\dfrac{(r^2-r_+^2)(r^2-r_-^2)}{r^2}d\tau_E^2+\dfrac{r^2dr^2}{(r^2-r_+^2)(r^2-r_-^2)}+r^2 \Big(dx+i\dfrac{r_+r_-}{r^2}d\tau \Big)^2\Big),
\ea
 where $x\sim x+2\pi$, $r_+\ge r_-$, and $l$ reflects the AdS radius. The rotating BTZ black hole has the following mass and angular momentum: 
 \ba\label{MAS2}
 8G_3M=r_+^2+r_-^2, \quad J=\dfrac{lr_+r_-}{4G_3},
 \ea
 where $r_+$ and $r_-$ are dimensionless in our convention. 
The spacetime is identified as~\cite{Carlip:1995qv} 
$\tau_E\sim \tau_E +\beta$ and $x\sim x-i\beta \Omega$, 
where the inverse temperature was determined by $\beta =2\pi r_+/(r_+^2-r_-^2)$ and the chemical potential as $\Omega =r_-/r_+$. The periodicity is characterized by the left and right temperatures following the addition of a complex coordinate $z=x+i \tau_E$: $\beta_L=\beta (1-\Omega)=\frac{2\pi}{r_++r_-}$ and $\beta_R=\beta (1+\Omega)=\frac{2\pi}{r_+-r_-}$.

The rotating BTZ black hole \eqref{BTZ1} is dual to the thermal CFT with finite temperature $\beta^{-1}$ and possesses a chemical potential $\Omega$ for momentum. 
Following the identification mentioned above, the dual CFT is described by the following density matrix: 
\ba\label{DEN1}
\rho= e^{-\beta H+\beta \Omega P}=e^{-\beta H+i\beta \Omega_E P},
\ea
 where the Hamiltonian and momentum are the components of it, and $\Omega_E=-i\Omega$ is an imaginary chemical potential. This density matrix meets the Hermitian condition $\rho^{\dagger}=\rho$ when $\Omega$ is real. 

Generally speaking, the double Wick rotated theory is described by the transition matrix rather than the density matrix. We use the double Wick rotation in the Euclidean signature to obtain the transition matrix of the double Wick rotated theory, namely, $z\to -iz'$, where $z'=\tilde{x}+i\tilde{\tau}_E$. $z'$ and $\bar{z}'$ have two different periodicities along spatial directions satisfying $L_1=-\beta_L$ and $L_2=-\beta_R$, respectively. Periodicity of $\tilde{\tau}_E$ is considered to be infinite: double Wick rotated theory describes CFT at zero temperature for real $r_-$. In terms of components of $z'=\tilde{x}+i\tilde{\tau}_E$, the periodicity is exchanged $\tilde{\tau}\to \tilde{\tau}-i\beta \Omega$ and  $\tilde{x}\sim \tilde{x}-\beta $.

{Thus, the transition matrix of the double Wick rotated theory is 
\begin{equation}
\rho' = e^{-i\beta \tilde P + i\beta \Omega \tilde H}=e^{-i\beta \tilde P -\beta \Omega_E \tilde H},
\label{DEN11}
\end{equation}
where $\tilde H$ and $\tilde P$ refer to the Hamiltonian and momentum of the double Wick rotated theory, respectively. $\tilde H$ should be interpreted as an analytic continuation $t\to -i\tilde{x}$ of the time-like Hamiltonian defined in \cite{Doi:2023zaf} and then become  a conventional Hamiltonian (see also section \ref{SEC6}). This ensemble correctly reproduces the periodicity mentioned above. In this sense, the parameter multiplying $\tilde P$ and $\tilde H$ may be viewed as an imaginary chemical potential and an (imaginary) inverse temperature, respectively.}

The ensemble of \eqref{DEN11} (the second equality) is in a more conventional form  at an imaginary value $\Omega=i\Omega_E$.  
The periodicity becomes 
$\tilde\tau_E \sim \tilde\tau_E + \beta\Omega_E$ 
and
$\tilde x \sim \tilde x - \beta$. Therefore, after exchanging the roles of the two cycles of the boundary torus, we define
\begin{equation}
\beta'=\beta\Omega_E,
\qquad
\beta=-\beta' \Omega_E' .
\label{PER12}
\end{equation}
With this identification, the same ensemble can be rewritten as
\begin{equation}
Z=\mbox{Tr}(\rho'),\quad \rho' = e^{-\beta' \tilde H + i\beta' \Omega_E' \tilde P}.
\label{DEN13}
\end{equation}
Equation \eqref{DEN13} should be understood as a formal re-quantization of the same torus after exchanging the thermal and spatial cycles, rather than as an ordinary thermal density matrix with a Hermitian Hamiltonian~\eqref{DEN1}. 

Due to diffeomorphism invariance, nevertheless, free energy $F=-\beta^{-1}\log Z$ matches the thermal CFT \eqref{DEN1} in the gravity dual calculation if we identify $\beta'$ and $\Omega_E'$ in \eqref{DEN13} with inverse temperature and an imaginary chemical potential~\cite{Dai:2025yhi}. Actually, this invariance is  ``without the double Wick rotation'' as shown in section \ref{SEC5}. Note that the modular transformation $\tau\to -1/\tau'$ of CFT dual to a $3d$ BTZ black hole is a different illustration of the double Wick rotation in string theory~\cite{Kraus:2006wn}, where the partition function is subject to changes in its periodicity and boundary conditions.   

By considering the conventional ensemble \eqref{DEN13} and tracing out in the subsystem $B$, one obtains the reduced transition matrix $\rho_A'=\mbox{tr}_B(\rho')$. The observer can only access subsystem A. The von-Neumann entropy in terms of $\rho_A'$ is the geometric entropy as follows:
\ba
S_G=-\mbox{Tr}(\rho_A'\log \rho_A').
\ea
The geometric entropy is different from the entanglement entropy. However, these are connected by the double Wick rotation~\cite{Fujita:2008zv}.

\section{Expectation values of stress tensor from CFT}\label{SEC3}
In this section, we derive expectation values of the stress tensor from the dual CFT. We define the complex coordinate $z=x+i\tau_E$. For a conformal transformation $z \mapsto f(z)= e^{i z/\tau}$ $(z \sim z + 2\pi)$ from the Euclidean cylinder to the plane, the quantum stress tensor satisfies
\begin{equation}
(\partial f)^2\, T(f) = T(z) - \frac{c}{12}\,\{f,z\},
\label{eq:T-transform}
\end{equation}
with the Schwarzian derivative
\begin{equation}
\{f,z\}
 = \frac{2 f' f''' - 3 f''^2}{2 f'^2}=\dfrac{1}{2\tau^2},
\label{eq:Schwarzian}
\end{equation}
{where $c$ is the central charge. Due to boundary conditions, the central charge serves as a response to macroscopic length scales, such as the circumference of a cylinder~\cite{DiFrancesco:1997nk}.}

Using the transformation formula from the cylinder (\ref{eq:T-transform}) and the vacuum condition $\langle T(f)\rangle = 0$ on the plane,
\begin{equation}
\langle T(z) \rangle = \frac{c}{24 \tau^2},
\qquad
\langle \bar T(\bar z) \rangle = \frac{c}{24 \bar{\tau}^2}.
\label{eq:T-cylinder}
\end{equation}
{Changes in energy density are caused by the periodic boundary condition, which is proportional to the central charge. The expectation value goes to zero in the large $\tau(\bar{\tau})$ limit (the large scale).}

\subsection{A rotating BTZ black hole}
The Euclidean BTZ black hole corresponds to a boundary torus with the modular parameter $\tau$. {The metric can be restricted to $\mbox{Im}(\tau )>0$ since it remains invariant under complex conjugation and degenerates for a real $\tau$.}  Using (\ref{eq:T-cylinder}), the Virasoro charges follow directly from $\tau$.

For the rotating BTZ black hole, we have the modular parameter $\tau = \frac{i\, l}{r_+ + r_-}$, where $r_-$ is pure imaginary and $\mbox{Im}(r_-)<0$. Using \eqref{eq:T-cylinder}, energy and momentum are obtained from the Virasoro zero modes:
\begin{equation}
M =\dfrac{1}{2\pi}\!\int_0^{2\pi}\! dx\, \langle T_{\tau_E\tau_E}(z)\rangle = \frac{r_+^2 + r_-^2}{8 G l^2},
\quad
P =\dfrac{1}{2\pi}\!\int_0^{2\pi}\! dx (\langle T(z) \rangle - \langle \bar{T}(\bar{z})) )= -\,\frac{r_+ r_-}{4 G l^2},
\end{equation}
where dimensions are recovered by $(r_+,r_-)\to (r_+/l,r_-/l)$ in \eqref{MAS2}. {Note that the expectation value of the stress tensor $\bar{T}(\bar{z})$ goes to zero in the extremal limit $r_+=r_-$ to recover the ground state at zero temperature in a right-moving sector.} 

\subsection{``DWR BTZ'' Identification}
The double Wick rotation exchanges two coordinates $x\leftrightarrow \tau_E$. Using the modular parameter $\tau = -\frac{l}{r_+ -r_-}$, similarly, energy and momentum are the same functional form as follows:
\begin{equation}\label{EPL22}
El = -\,\frac{r_+^2 + r_-^2}{8 G l},
\qquad
Pl = -\,\frac{r_+ r_-}{4G l}.
\end{equation}
{The expectation value yields a vacuum energy density of nonzero on the Euclidean cylinder for the double Wick rotated (DWR) identification and agrees with the gravity dual~\cite{Fujita:2022zvz}.}




\section{The double Wick rotated BTZ black hole as quotients}\label{SEC4}
This section provides us with the double Wick rotated BTZ black hole that is quotients in $AdS_3$. The Euclidean $AdS_3$ metric is written as the coset space ${SL}(2,C)/SU(2)$~\cite{Carlip:1994gc}. Dimensions of these are $6-3=3$ as expected for 3 dimensional spacetime. The isometry of $SL(2,C)$ is manifest in the coset space. After performing identifications, the thermal AdS is obtained by quotienting $AdS_3$. 

{We show that the double Wick rotated metric is in a $SL(2,\mathbb{Z})$ family of black holes for an imaginary $r_-$ under the modular transformation.} That is, there is the modular transformation between the thermal AdS with a modular parameter $\tau$ with the double Wick rotated {background} with a modular parameter $\tau'=-1/\tau$. Because the double Wick rotated metric is obtained by $z\to iz'$ in a rotating BTZ black hole \eqref{BTZ1}, the modular parameter is $\tau'=l/(r_++r_-)$ with real $r_+$ and pure imaginary $r_-$ ($\mbox{Im}(r_-)<0$) compared with the BTZ modular parameter $\tau'=il/(r_++r_-)$ \cite{Carlip:1994gc,Kraus:2006wn}),  
 The modular transformation involves a coordinate transformation in the bulk~\cite{Kraus:2006wn}. The action, which is $-\log Z$ in the saddle point, becomes invariant under diffeomorphism from the thermal AdS to the double Wick rotated BTZ black hole as follows:
\ba\label{ACT47}
\log Z=-\dfrac{i\pi}{12}(c\tau-\bar{c}\bar{\tau})=\dfrac{i\pi}{12}\Big(\dfrac{c}{\tau'}-\dfrac{\bar{c}}{\bar{\tau'}}\Big)+\dots 
\ea
In the high temperature limit, $\mbox{Im}(1/\tau')\to -\infty$, the double Wick rotated BTZ black hole becomes more stable than the thermal AdS.

 According to~\cite{Fujita:2022zvz}, the double Wick rotated metric is locally equivalent to the pure $AdS_3$ $ds^2=l^2\frac{d\rho^2+dWd\bar{W}}{\rho^2}$. The double Wick rotated BTZ black hole is obtained as follows:
\ba\label{DWR41}
ds^2=l^2\Big(r^2\Big(d\tau_E-\dfrac{ir_-r_+}{r^2}dx\Big)^2+\dfrac{(r^2-r_+^2)(r^2-r_-^2)}{r^2}dx^2+\dfrac{r^2dr^2}{(r^2-r_+^2)(r^2-r_-^2)} \Big),
\ea
where  the identification is given by $w_1'\backsimeq w_1'+2\pi \tau'\backsimeq w_1'+2\pi $. The metric \eqref{DWR41} is real metric and a black hole by performing an analytic continuation $\tilde{r}_-=-ir_-$. The metric \eqref{DWR41} is the double Wick rotation of \eqref{BTZ1}: $\tau_E\leftrightarrow -x$. Following~\cite{Kraus:2006wn}, the Virasoro zero modes are obtained from \eqref{ACT47}. It agrees with \eqref{EPL22}. The entropy (the Cardy formula) with the Virasoro zero modes is nonzero for imaginary $r_-$.

\section{Equivalence of the double Wick rotated BTZ black hole}\label{SEC5}
In this section, we compare the observables of a rotating BTZ black hole and a double Wick-rotated BTZ black hole with the same periodicity. Actually, two black holes are related to each other by a coordinate transformation without the double Wick rotation. This equivalence can be used to determine observables such as the boundary stress tensor and the entanglement entropy in the double Wick-rotated background in the Lorentzian signature and at angular momentum. 

We account for the Lorentzian signature. After the analytic continuation $\tau_E=-it$, the double Wick rotated BTZ \eqref{DWR41} has the following ADM form:
\ba\label{DWR42}
ds^2=l^2\Big(-\dfrac{(r^2-r_-^2)(r^2-r_+^2)}{r^2-r_-^2-r_+^2}dt^2+(r^2-r_-^2-r_+^2)\Big(dx-\dfrac{r_+r_-dt}{r^2-r_-^2-r_+^2}\Big)^2+\dfrac{r^2dr^2}{(r^2-r_+^2)(r^2-r_-^2)}  \Big).
\ea
 Note that the boundary of the ergosphere $r_g^2=r_+^2+r_-^2$ in a rotating BTZ black hole is mapped to the boundary of closed timelike curves.

Exchanging the parameter $r_+\leftrightarrow \tilde{r}_-(=-ir_-)$ in \eqref{DWR42} leads to the following black hole solution:
\ba\label{DWR2}
ds^2=l^2\Big(-\dfrac{(r^2-\tilde{r}_-^2)(r^2+r_+^2)}{r^2-\tilde{r}_-^2+r_+^2}dt^2+(r^2-\tilde{r}_-^2+r_+^2)\Big(dx-\dfrac{ir_+\tilde{r}_-dt}{r^2-\tilde{r}_-^2+r_+^2}\Big)^2+\dfrac{r^2dr^2}{(r^2+r_+^2)(r^2-\tilde{r}_-^2)}  \Big),
\ea  
The black hole horizon is located at $r=\tilde{r}_-$. The metric \eqref{DWR2} with the Euclidean signature is a Riemannian manifold and positive definite: $g_{mn}=\delta_{\alpha\beta}e^{\alpha}_me^{\beta}_n$ and $g_{mn}e^{m}_{\alpha}e^n_{\beta}=\delta_{\alpha\beta}$, where we have defined a tangent frame field $e^{m}_{\alpha}$ and a Kronbecker symbol  $\delta_{\alpha\beta}$. The periodicity of the torus agrees between two black holes as pointed out in~\cite{Dai:2025yhi}. Performing the following coordinate transformation
\ba\label{COO55}
r^2=u^2-r_+^2+\tilde{r}^2_-,
\ea
the metric is rewritten as a rotating BTZ metric \eqref{BTZ1} with $\tilde{r}_-$. As expected, the black hole horizon is mapped to $u=r_+$. 

In accordance with \cite{Fujita:2022zvz} and utilizing the covariance between \eqref{BTZ1} and \eqref{DWR2}, therefore, observables in the double Wick rotated metric \eqref{DWR42} will be obtained by exchanging $r_+$ for $\tilde{r}_-$ for those in a rotating BTZ as follows:
\ba\label{OBZ57}
O^{DWR}(r_+,\tilde{r}_-)=O^{BTZ}(\tilde{r}_-,r_+)\quad \mbox{or} \quad O^{DWR}(\beta,\beta\Omega_E )=O^{BTZ}(\beta\Omega_E , \beta).
\ea
One can confirm \eqref{OBZ57} from the expectation values of the boundary stress tensor $T_{\mu\nu}$~\cite{He:2014lfa,Fujita:2022zvz}, thermodynamics, total energy of spacetime, and holographic two-point functions~\cite{Dai:2025yhi}. 

The entanglement entropy with a double Wick rotation is known as geometric entropy. We take into account the holographic entanglement entropy in the general interval~\cite{Hubeny:2007xt} as follows:
\ba\label{EEN33}
S_A=\dfrac{c}{6}\log \Big(\dfrac{\beta^2(1+\Omega_E^2)}{\pi^2\epsilon^2}\sinh \Big(\dfrac{\pi \Delta w'}{\beta (1+i\Omega_E)} \Big)\sinh \Big(\dfrac{\pi \Delta \bar{w}'}{\beta (1-i\Omega_E)} \Big) \Big),
\ea
where $\Delta w'= w_1'-w_2'$. The entanglement entropy in a rotating BTZ black hole is linked to the geometric entropy by exchanging $r_+$ and $\tilde{r}_-$, namely, exchanges of $\beta$ and $\beta\Omega_E$. These exchanges transform sinh functions into sin functions. Finally, we obtain 
\ba\label{GEO34}
S_G=\dfrac{c}{6}\log \Big(\dfrac{\beta^2(1+\Omega_E^2)}{\pi^2\epsilon^2}\sin \Big(\dfrac{\pi \Delta w'}{\beta (1-i\Omega_E)} \Big)\sin \Big(\dfrac{\pi \Delta \bar{w}'}{\beta (1+i\Omega_E)} \Big) \Big).
\ea
\eqref{GEO34} agrees with the result in~\cite{Fujita:2022zvz} when $\Delta w'$ is the spatial length of the interval. The geometric entropy in \eqref{GEO34} is also obtained via the double Wick rotation $w'\to i\bar{z'}$ of \eqref{EEN33}.~\footnote{The complex conjugation depends on the sign of periodicity. When the identification \eqref{PER12} is considered, there are no complex conjugation.}



\section{Time-like entanglement entropy in the rotating BTZ black hole}\label{SEC6}



In this section, we derive the time-like entanglement entropy from an
appropriate analytic identification and analysis of its time dependence.
For time-like entanglement entropy, we consider the transformation
$x=i\tilde{t}$ and $t=\tilde{x}$. We then obtain the time-like Hamiltonian
\ba\label{TRA1}
P=-i\partial_x =-\partial_{\tilde{t}}=i\tilde{H},\quad
H=i\partial_t=-i\partial_{\tilde{x}}=-\tilde{P},
\ea
where $\tilde{H}$ is an anti-Hermitian timelike Hamiltonian, while
$\tilde{P}$ remains Hermitian.
We further perform a transformation that maps the generator to a Hermitian
operator,
\ba\label{HAM36}
\tilde{H}=-i\hat{H},
\ea
so that $\hat{H}$ takes the form of a conventional Hamiltonian.
For a free scalar theory, this corresponds to the analytic continuation
$m\to im$.
We introduce a complex moduli parameter
$\tau=\beta\Omega_E+i\beta$ describing a thermal CFT.
Using~\eqref{TRA1}, the density matrix is transformed into the transition matrix
\ba
\rho''=e^{\beta\tilde{P}-\beta\Omega_E\tilde{H}}
      =e^{\beta P+i\beta\Omega_E\hat{H}} .
\ea

The appearance of the complex evolution operator admits a natural
holographic interpretation. The analytic continuation leading to
(\ref{TRA1}) effectively exchanges spacelike and timelike directions and
induces a complex identification of the rotating BTZ geometry.
From the bulk perspective, this continuation corresponds to evolving
along a complexified modular direction rather than an ordinary unitary
time evolution. In semiclassical AdS/CFT, such analytically continued
configurations are expected to be described by complex saddles of the
gravitational path integral \cite{Maldacena:2001kr}. The resulting
non-Hermitian structure therefore reflects the analytic structure of the
dual geometry rather than a breakdown of unitarity. In this picture, the
imaginary parts of the spectrum encode phases generated by Lorentzian
modular flow, consistent with recent developments in time-like
entanglement and pseudo-entropy \cite{Doi:2022iyj,Doi:2023zaf}.
The moduli parameter is transformed as
\ba\label{TRA2}
\tau_1'=R=-i\beta',\quad
\tau_2'=\beta=-i\beta'\Omega_E'.
\ea
Compared with \eqref{PER12}, it is multiplied by an imaginary number. This factor arises from an analytic continuation $t\to -i\tilde{x}$.

We now consider the entanglement entropy for the general interval
\eqref{EEN33}. After performing the transformation~\eqref{TRA2},
$\Delta w'=\Delta \tilde{t}$ and $\epsilon=-i\tilde{\epsilon}$, one obtains
the time-like entanglement entropy
\ba\label{TIM39}
S_{\mathrm{TL}}
=\dfrac{c}{6}\log \Big[
\dfrac{\beta^{\prime 2}(1+\Omega_E^{\prime 2})}{\pi^2\epsilon^{\prime 2}}
\sinh \Big(\dfrac{\pi\Delta \tilde{t}}{\beta'(1+i\Omega_E')}\Big)
\sinh \Big(\dfrac{\pi\Delta \tilde{t}}{\beta'(1-i\Omega_E')}\Big)
\Big]
+\dfrac{\pi ic}{6},
\ea
where the energy scale $\epsilon$ is analytically continued due to
\eqref{HAM36}. Equation~\eqref{TIM39} exactly agrees with the result obtained in~\cite{Doi:2023zaf}.
The time-like entanglement entropy grows linearly at late times,
\begin{equation}
S_{\mathrm{TL}}(t \to \infty)
\simeq
\dfrac{c\pi t}{3\beta'(1+\Omega_E'^2)}
=
\frac{c}{6} r_+ t ,
\label{eq:LinearGrowth_main}
\end{equation}
suggesting a universal late-time growth behavior governed by the outer
horizon radius.
Motivated by this behavior, we propose that the coefficient of linear
growth defines a Lorentzian entanglement growth exponent,
\begin{equation}
\lambda_{\mathrm{TL}}
=\frac{c}{6} r_+ .
\label{eq:lambdaTL_main}
\end{equation}

In contrast to the usual Lyapunov exponent
$\lambda_L=2\pi T$ extracted from out-of-time-ordered correlators, which vanishes in the extremal limit $T\to0$, the quantity
$\lambda_{\mathrm{TL}}$ remains finite.
Time-like entanglement entropy therefore provides a complementary diagnostic of holographic chaos that survives at extremality. It probes the rate at which Lorentzian correlations propagate along the boundary time direction and geometrically reflects the connectivity induced by the near-horizon region. Its persistence in the extremal limit suggests that it captures dynamical information beyond conventional thermal chaos indicators.

\section{Summary}
In this paper, we analyzed geometric entropy and time-like entanglement entropy in a rotating BTZ black hole in the holographic context . The transition matrix  of double Wick rotated theory was developed, which has the typical shape at an imaginary chemical potential after exchanging periodicity \eqref{PER12}. 
{When combined with the BTZ modular parameter, the Schwarzian derivative \eqref{eq:Schwarzian} in a conformal transformation from the cylinder helps us understand the Virasoro zero modes $(L_0,\bar L_0) \quad \Longleftrightarrow \quad (E,J)$ via the AdS$_3$/CFT$_2$ correspondence.}

Exchanging periodicity corresponds to the equivalence between a rotating BTZ black hole and the double Wick rotated metric in Section \ref{SEC5}. {We showed that the double Wick rotated metric was equal to a rotating BTZ black hole after the coordinate transformation \eqref{COO55} if periodicity is identified. Due to this equivalence and the identification \eqref{OBZ57}, the geometric entropy in the general interval was derived in \eqref{GEO34}, which was the double Wick rotation of the entanglement entropy. Thus, this identification will be helpful to determine observables (Choptuik scaling~\cite{Birmingham:2001hc} etc.) obtained in the double Wick rotated theory at a chemical potential. 


Time-like entanglement entropy is obtained by using the identification, which generalizes the identification in \cite{Doi:2023zaf} for the general moduli parameter. We find a new Lorentzian entanglement-growth exponent from the coefficient of the linear growth of time-like entanglement entropy. It will be interesting to extend the analysis in this paper to $2d$ CFT on the torus with the general moduli parameter or a massive fermion.}

\section*{Acknowledgments}
We thank Wuzhong Guo, Jia-Rui Sun, and Tadashi Takayanagi for valuable discussion and comments. This work was supported by NSFC Grant Nos. 12475053, 12588101, and 12235016, and the sub-project funding for “Gravitational Redshift Measurement Scientific Experiment and Frontier Research in Gravitational Physics” of the Chinese Academy of Sciences, the Strategic Priority Research Program on Space Science, the Chinese Academy of Sciences (XDA30040000, XDA30030000).

\end{document}